\documentclass[12pt]{article}
\usepackage{latexsym}

\newcommand{\beq}{\begin{equation}}
\newcommand{\eeq}[1]{\label{#1}\end{equation}}
\newcommand{\bea}{\begin{eqnarray}}
\newcommand{\eea}[1]{\label{#1}\end{eqnarray}}

\begin{document}
\begin{titlepage}
\hfill NYU-TH/01/08/02 hep-th/0109017
\vspace{20pt}

\begin{center}
{\large\bf{MASS AND GAUGE INVARIANCE IV \\ (HOLOGRAPHY FOR THE KARCH-RANDALL 
MODEL)}}
\end{center}

\vspace{6pt}

\begin{center}
{\large M. Porrati} \vspace{20pt}

{\em Department of Physics, NYU, 4 Washington Pl, New York NY 10003}

\end{center}

\vspace{12pt}

\begin{center}
\textbf{Abstract }
\end{center}
\begin{quotation}\noindent
We argue that the Karch-Randall compactification is holographically 
dual to a 4-d conformal field theory coupled to gravity 
on Anti de Sitter space.
Using this interpretation we recover the mass spectrum of the  
model. In particular, we find no massless spin-2 states.
By giving a purely 4-d interpretation to the compactification
we make clear that it represents the first example of a local 4-d 
field theory in which general covariance does not imply the existence of a 
massless graviton. We also discuss some variations of the Karch-Randall
model discussed in the literature, and we examine whether its properties are 
generic to all conformal field theory.
\end{quotation}
\vfill
\end{titlepage} 
\section{Introduction}
\noindent
Nowadays, it is commonly known that gauge invariance does not guarantee the
existence of massless spin-one particles. The W and Z particles of the 
Standard Model are the best known example of this fact, but it was noticed
first by Julian Schwinger in his seminal article~\cite{s} 
echoed by our title.

General covariance is a different story. Until recently no example was known 
of a theory invariant under general coordinate transformations in four 
dimensions that did not contain a massless spin-2 particle. String theory
and Kaluza-Klein compactifications contain many massive states of 
spin-2 (and higher), but they always have a unique massless graviton.

The first example of a consistent generally covariant theory without {\em any} 
massless spin-2 state has been given only recently by Karch and 
Randall~\cite{kr}, who studied a compactification similar to the  
Randall-Sundrum II model (RSII). In RSII~\cite{rs2}, 
pure 5-d Einstein gravity with a 
negative cosmological constant is ``compactified'' on an Anti de Sitter (AdS) 
space cut off near its boundary by a flat 4-d brane, i.e. a brane with flat
induced metric. In spite of the fact that the regularized AdS space has an 
unbounded transverse coordinate, RSII is a compactification, since the 5-d 
graviton has a normalizable zero mode, localized near the 4-d brane, that 
gives rise to the standard 4-d Einstein gravity up to small corrections, due
to the exchange of massive spin-2 states. 

The Karch-Randall model (KR) is a variation of RSII, in that the induced 
metric on the 4-d brane is itself AdS. The 4-d cosmological constant of the 
induced metric is an additional free parameter, besides the 5-d cosmological
constant previously mentioned. The 4-d cosmological constant can be set to
zero, and the KR model is continuous in that limit. 
In our opinion, the greatest surprise of the KR compactification is that it
has {\em no} massless spin-2 states; instead, its spectrum consists of 
a tower of massive spin-2 states, of mass $O(\lambda)$, and a lighter state
of mass $O(\lambda^2/M_{Pl}^2)$. Here $\lambda$ is the 4-d ``cosmologist's''
cosmological constant, i.e. the vacuum energy $V$ divided by the 4-d 
Planck constant: $\lambda=V/M_{Pl}^2$. Yet, this model {\em is} generally
covariant in 4-d, as shown in~\cite{kr} and more extensively in~\cite{kkr}.
Thus, the KR model is the first example in which general covariance does not
imply the existence of a massless spin-2 particle.

Before proceeding further, let us solve an apparent puzzle. In the limit
$\lambda\rightarrow 0$, the KR model goes smoothly into the RSII model, which 
does possess a massless graviton. The smoothness of the limit is not in 
contradiction with the well-known van Dam-Veltman-Zakharov (vDVZ)
discontinuity~\cite{vvdz}, or other discontinuities, for two reasons.
First of all, the vDVZ discontinuity does not exist in AdS space. More 
accurately, the tree-level, one-particle exchange amplitude 
of a massive spin-2 field in 4-d AdS space is continuous in the limit
$m^2/\lambda\rightarrow 0$~\cite{p} (see also~\cite{kmp}). 
A discontinuity in quantum loops involving the massive graviton was claimed
to exist in ref.~\cite{dls}. Ref.~\cite{dls} works in a theory with a 
single massive spin-2, so its results may not apply to KR, 
which has a much richer spectrum. Anyway, that discontinuity 
is not relevant here, as throughout our paper we work in the weak-gravity 
regime, where the semiclassical approximation for 4-d gravity is valid. In 
that approximation all loops involving the 4-d graviton, 
hence their discontinuities, are negligible.
Finally, KR may be altogether free from quantum discontinuities 
as it may be realizable as a string theory background,
continuous in the limit $\lambda\rightarrow 0$, as argued
in~\cite{kr2}.

Randall-Sundrum compactifications admit a dual interpretation as 
four-dimensional local field theories coupled to 
gravity~\cite{Ed2,ahpr,prz}. 
In the dual theory, the field theory is strongly interacting, but gravity is 
weakly interacting at all scales up to the cutoff, which 
must be, therefore, smaller than the 4-d Planck scale.  
This duality is a consequence of the holographic duality~\cite{malda,gkp,Ed}, 
as most 
clearly pointed out in~\cite{hhr}. The arguments used in~\cite{hhr} to
justify the duality between RS compactifications and 4-d field theories 
coupled to standard gravity, hereafter called 
``gauged holography,''~\cite{ahpr} do not require that the 
metric on the 4-d brane is flat.
Indeed, the brane metric used in~\cite{hhr} is 
that of a Eucidean 4-sphere.
If gauged holography works for curved branes as well as Minkowsky branes,
we are naturally led to conjecture that even the KR compactification 
must admit a holographic dual. Clearly stated, we propose that\vskip .1in
{\em The Karch-Randall compactification is dual to a four-dimensional 
conformal field theory 
on a four-dimensional AdS space, coupled to (regularized) Einstein gravity}   
\vskip .1in
As demanded by the holographic duality, the 4-d 
conformal field theory (CFT) is strongly interacting when 
the five-dimensional description is semiclassical. Four-dimensional gravity
is weak.
The rest of this paper is devoted to prove the claim made here. 

In section 2, after briefly reviewing the KR compactification, we 
adapt and expand the treatment of gauged holography
given in~\cite{ahpr,hhr} to cover the case of arbitrary 4-d induced
metrics on the brane. We exhibit a very simple argument showing that
gauged holography, both for Minkowsky 4-d metrics and for curved 4-d metrics,
is not an independent conjecture, but that it follows 
instead from the ``rigid'' holographic duality. We also specialize the
discussion of gauged holography to the case that the metric on
the brane is a small fluctuation around a conformally flat background.
In that case, we will be able to show that if a RS compactification
has a holographic dual, then the KR compactification related to it
{\em must} have a holographic dual too, and that the dual CFT  
is the same in both cases.  

Section 3 treats the ``universal'' part of the effective action of a CFT on
a curved background, namely the term that comes from integrating the
Weyl anomaly: the Riegert action~\cite{r}. 
Unlike two-dimensional CFTs, CFTs in four dimensions
cannot be solved completely by integrating the Weyl anomaly, since 4-d
metrics admit infinitely many different conformal classes. This implies 
that the effective action of a 4-d CFT contains a model-dependent, 
Weyl-invariant piece, besides the Riegert action. In section 3, we will show 
that the Riegert action does not give a mass to the graviton~\footnote{The
Riegert action is ill-defined in the infrared. In this paper, that name always
denotes an appropriately regularized version of that action.}. The graviton 
mass found in the KR compactification is thus a highly non-trivial effect
due to the model-dependent term in the effective action. This term, it 
is worth repeating, cannot be determined by Weyl invariance alone.

Section 4 uses the holographic duality to compute
the effective action of gravity coupled to a CFT on an AdS background, 
expanded to quadratic order in the metric fluctuations. This is the term
that gives the two-point correlator of the stress-energy tensor. In section
4 we show that the same term also gives a mass $O(\lambda^2)$ to the 4-d 
graviton. Finally, section 4 shows that the two-point correlator so computed
has the correct flat-space limit, proportional to 
$p^4 \log p^2/\mu^2$~\cite{gkp}.

Section 5 contains our conclusions and a brief discussion of whether other 
behaviors for the two-point function of the graviton are possible for generic 
CFTs on AdS spaces. We also discuss the case, studied in~\cite{kmp2}, of 5-d 
AdS space bound by {\em two} positive-tension AdS branes. In particular,
we show how, in 
that case, the holographic duality implies immediately that a massless graviton
must exist, as shown in~\cite{kmp2}.   

A discussion of Weyl transformations, diffeomorphisms, and the counting of
degrees of freedom for the KR model is given in appendix A. Appendix B
presents an explicit, simple 
change of coordinates for $\mbox{AdS}_d$ that maps its 
Poincar\'e coordinates into new coordinates in which $\mbox{AdS}_d$ 
is sliced by $AdS_{d-1}$ surfaces.
\section{KR and Gauged Holography in Curved Space}
\subsection{KR (Abridged)}
To describe the Karch-Randall compactification~\cite{kr}, let us 
starts with the Einstein-Hilbert action in five 
dimensions, with a negative 5-d cosmological constant $\Lambda$,
\beq
S_{EH}={1\over 16\pi G}\int d^5x \sqrt{-g} (R -2\Lambda).
\eeq{1}
A solution of the Einstein equations derived from this action is the 
Anti de Sitter space. Among its many equivalent metrics, we choose one in 
terms of a space-like radial coordinate $z$, 
ranging from $-\pi L/2 $ to $+\pi L/2$ and four
other coordinates $x^\mu$, $\mu=0,1,2,3$~\footnote{Appendix B exhibits a 
reparametrization that transforms the Poincar\'e coordinates into those
of eq.~(\ref{2}).}.
\beq
ds^2={1\over \cos^2 (z/L)}(dz^2+ds_4^2), \qquad \Lambda\equiv -6/L^2.
\eeq{2}
The 4-d section, with line element $ds^2_4$, is an $\mbox{AdS}_4$ space with
radius $L$. 

The AdS space can be truncated by restricting $z$ to the range $-\pi L/2 \leq 
z \leq +\pi L/2 -\epsilon$, $\epsilon\ll L$, 
together with appropriate (e.g. Neumann) boundary
conditions at $z=+\pi L/2 -\epsilon$. This is the Karch-Randall 
compactification\footnote{A physical but by no means unique way to truncate 
the space is to place a 4-d brane of appropriate tension at the boundary.}.
Notice that the induced metric on the boundary is $ds^2=
\sin^{-2}(\epsilon/L)ds_4^2$. Because of the scaling factor in front of
$ds_4^2$, the induced metric has a negative
4-d cosmological constant $\lambda=-3\sin^2(\epsilon/L)/L^2
\approx -3\epsilon^2/L^4$.
The 4-d cosmological constant vanishes with $\epsilon$, and as expected the 
RS compactification is indeed recovered in the limit $\epsilon\rightarrow 0$. 
In particular, with the change of variables 
$\hat{z}=(L/\epsilon)(\pi L/2 -z)$, $\hat{x}^\mu
=(L/\epsilon)x^\mu$, the 5-d metric assumes in the limit the standard RS form 
(see for instance~\cite{ahpr})
\beq
ds^2= 
{L^2\over\hat{z}^2}(d\hat{z}^2 +\eta_{\mu\nu} d\hat{x}^\mu d\hat{x}^\nu),
\qquad \hat{z}\geq L.
\eeq{3} 
\subsection{Rigid and Gauged Holography I}
The holographic duality states that the generating functional of a 4-d 
conformal field theory is the partition function of quantum gravity on 
a 5-d manifold $X$. In particular, given a 5-d field $\phi(x,z)$, with 5-d 
1PI action $\Gamma[\phi]$, its boundary 
value $\phi(x,0)$ is the source for a gauge-invariant operator $O$ 
in the CFT, and the partition function is given by
\beq
\langle \exp[-\int_M d^4 x \phi(x) O(x)] \rangle_{CFT} =
\exp(-\Gamma[\phi]). 
\eeq{4} 
The manifold $X$ is a solution of the 5-d Einstein equations with 
cosmological constant $\Lambda$. $M$, the boundary of $X$, 
is the space on which the CFT lives,
and near the boundary $z=0$ the $X$ metric is
\bea
ds^2 &=& {L^2\over z^2}[dz^2 + g_{\mu\nu}(x,z)dx^\mu dx^\nu], \label{5}\\
g_{\mu\nu}(x,z)&=&g^0_{\mu\nu}(x) + z^2 g^1_{\mu\nu}(x) + z^4\log z^2 
g^2_{\mu\nu}(x) + O(z^4).  
\eea{6}
The metric on $M$ is 
$g^0_{\mu\nu}(x)$ and both $g^1_{\mu\nu}(x)$ and $g^2_{\mu\nu}(x)$ are
local function of $g^0_{\mu\nu}(x)$ and its derivatives.
Eqs.~(\ref{5},\ref{6}) define the metric up to diffeomorphism that act
on $g^0_{\mu\nu}(x)$ as conformal transformations. This property will be 
very useful later on.

To compute $\Gamma[\phi]$, we use the semiclassical approximation, where
$\Gamma[\phi]$ is the classical action {\em on shell}. Even in this
approximation $\Gamma[\phi]$ must be regularized.

Let us consider in particular the case where $\phi$ is the 5-d metric, and
 $\Gamma[\phi]$ is the 5-d action of pure gravity with negative cosmological
constant.
The action is $\Gamma= S_{EH} + S_{GH}$. The Einstein-Hilber action 
$S_{EH}$ has been given in eq.~(\ref{1}), while the Gibbons-Hawking boundary
term, necessary to have an action that depends only on the first derivative 
of the metric~\cite{GH} is
\beq
S_{GH}= {1\over 8\pi G} \int d^4 x\sqrt{-h} K.
\eeq{7} 
$h$ is the determinant of the induced metric on the boundary and $K$ 
is the trace of the extrinsic curvature of the boundary. 
When computed on-shell, the action $S_{EH}+S_{GH}$ diverges. For instance,
the Einstein-Hilbert piece is 
\beq
S_{EH}={1\over 16\pi G}\int_0^\infty dz \int d^4x {L^5\over z^5} 
\left[{4\Lambda\over 3}\sqrt{-g^0(x)} +O(z^2)\right]=\infty.
\eeq{8}
To regularize it the integral in $z$ is cut off at some positive value 
$\epsilon$. The regularized action $\Gamma_\epsilon[g^0_{\mu\nu}(x)]$ has the 
following expansion in powers of $\epsilon$~\cite{hs,dhss}
\beq
\Gamma_\epsilon=\epsilon^{-4} a_0 +  \epsilon^{-2} a_2 + \log(\epsilon^2)a_4 +
\Gamma_\epsilon^F.
\eeq{9}
In this equation, $a_0$ is proportional to the 4-d cosmological constant term,
$\int_M d^4x \sqrt{-g^0}$, $a_2$ is proportional to the 4-d Einstein-Hilbert 
action, $\int_M d^4x\sqrt{-g^0} R(g^0)$, and $a_4$ is a linear combination of
the Euler curvature and the square of the Weyl tensor 
\beq
a_4={L^3\over 128\pi G} \int d^4 x
\sqrt{-g^0}\left ( {1\over 3}R^2(g^0)- R_{\mu\nu}(g^0)R^{\mu\nu}(g^0) 
\right).
\eeq{10}

The holographic duality can be resumed in one equation now, namely:
\beq
\lim_{\epsilon\rightarrow 0} \Gamma_\epsilon^F = W_{CFT}[g].
\eeq{11}
Here $W_{CFT}[g]$ is the generating functional of the (connected) 
correlators of the stress-energy tensor. 
Eq.~(\ref{11}) makes explicit a point that is often hidden in the
literature on the holographic duality. Namely, that the regularized
generating functional of the CFT is still given by a 5-d holographic dual
even when the cutoff $\epsilon$ is small but nonzero. 
Notice that when $g_{\mu\nu}$ is
expanded around flat space, $g_{\mu\nu}=\eta_{\mu\nu} + h_{\mu\nu}$, one 
obtains the ``usual'' correlators of $T_{\mu\nu}$ in Minkowsky space. On the 
other hand, we could have expanded eq.~(\ref{11}) around any background.
When expanded around a curved background, $W_{CFT}$ gives the stress-energy 
tensor correlators of the CFT on that curved background. This fact has 
been used time and again in the literature on the holographic duality.
After all, ref.~\cite{Ed} already studies holography not only on $R^4$ but
also on $S_4$ and $S_3\times S_1$.
The local, divergent terms $a_0,a_2,a_4$ are harmless in the standard, 
``rigid'' holographic duality, since they give rise only to contact terms. 

In ``rigid'' holography, the boundary value of the 5-d metric, $g^0_{\mu\nu}$,
is simply an external, fixed quantity, that acts as the source for the 
stress-energy tensor. On the other hand, 
holography was at first derived in string theory, where
the graviton is dynamical. This is not in contradiction with the 
previous assumption that $g^0_{\mu\nu}$ is an external source, as long as the
4-d Newton constant vanishes. This is the case in the limit $\epsilon
\rightarrow 0$, since eq.~(\ref{9}) implies $G_4\propto G\epsilon^2$. In 
other words, the induced 4-d Newton constant vanishes when the cutoff is 
removed. This result is consistent with the holographic interpretation. 
After all, in a 4-d CFT with ultraviolet cutoff $\epsilon$ the induced Newton
constant is indeed proportional to $\epsilon^2$.

If $\epsilon$ is kept finite, then it is natural to promote $g^0_{\mu\nu}$ to
a dynamical field. For instance, 
in string theory, when deriving the holographic
duality using, say, $D3$ branes, $g^0_{\mu\nu}$ is just the
value of a dynamical field at some arbitrary boundary in between an AdS 
``throat'' and an asymptotically flat region in the string background. In 
that case, the background is found by solving the equations of motion of
$g^0_{\mu\nu}$.
Moreover, it is natural, and allowed by the symmetries of our system, 
to add to the action $\Gamma_\epsilon$ local 4-d terms, proportional to
$a_0$, $a_2$, and $a_4$. The coefficient of the $a_0$ term 
is the best known in the literature as 
it is the tension of the Randall-Sundrum ``Planck'' brane~\cite{rs2,ahpr}.
It acts as a conterterm that cancels part of 
the induced cosmological constant in eq.~(\ref{9}).
The term proportional to $a_2$ 
was discussed in~\cite{ahpr} and its interpretation
is reviewed in the next paragraph. The third term is largely 
irrelevant when studying 4-d gravity at low energy.
It can be set equal to $-\log (\epsilon^2)a_4$ to simplify eq.~(\ref{9}),
without loss of generality.

Let us examine again the conclusion of the previous paragraph. We noticed 
that if $\epsilon$ is nonzero, and if we assume the standard ``rigid'' 
holographic duality, the on-shell 5-d Einstein action with cosmological 
constant $\Lambda=-6/L^2$, computed on the manifold $X$ with boundary condition
$g^4_{\mu\nu}\equiv (L/\epsilon)^2 g^0_{\mu\nu}$ on $\partial X=M$ is
\bea
\Gamma_H[g^4_{\mu\nu}]&=& {1\over 16\pi G_4}
\int_M d^4 x \sqrt{-g^4}( R-2\lambda) + W_{CFT}[g^4_{\mu\nu}] + ...,
\label{12}\\
{1\over 16\pi G_4}&=&{L\over 32\pi G}+ {1\over 16\pi G^{bare}_4},
\qquad {1\over 16\pi G_4}\lambda = -{3\Lambda L\over 16\pi G }-T.
\eea{13}
$T$ is the coefficient of
the $a_0$ term; we have already interpreted it as the tension of
a brane placed at the boundary~\footnote{The ``brane'' can be sometimes 
just an effective description of a more complicated mechanism see~\cite{v} 
for an early example of this possibility, and~\cite{gkp2} for a 
recent discussion.}. ${1\over 16\pi G^{bare}_4}$ is the coefficient of the 
$a_2$ term. As implied by our label, it does appear as a bare Newton
constant. The other coefficients given in eq.~(\ref{13}) are those of 
eq.~(\ref{9}); they have been computed, for
instance, in~\cite{dhss,dhss2}.

Notice that eq.~(\ref{12}) is {\em exactly} the effective action of gravity
coupled to a CFT with ultraviolet cutoff $L$, 
up to terms vanishing with the cutoff. Eq.~(\ref{12}) proves by itself 
that ``rigid'' 
holography implies, without any further assumption, 
``gauged'' holography, namely holography in the presence of dynamical gravity.

A few comments are now appropriate.
\begin{enumerate}
\item We have rescaled the boundary metric in eq.~(\ref{12}) so
that at $z=\epsilon$, $ds^2=g^4_{\mu\nu}dx^\mu dx^\nu$. In this fashion, the 
4-d length is given by the metric $g^4_{\mu\nu}$ without the need of any 
further rescaling. The definition of $g^4_{\mu\nu}$ also makes evident that
we can always set the cutoff at $L$, instead of $\epsilon$, by appropriately
rescaling the Newton and cosmological constants of the 4-d theory. 
\item Eq.~(\ref{12}) is evidently the effective action obtained after
integrating out the CFT, but before taking into account graviton loops.
That is the appropriate effective action whenever the CFT is strongly 
interacting but all true quantum gravity effects are negligible, i.e. when the
CFT is cut off at a scale well below both the 5-d and 4-d Planck scales.
The condition that the cut-off is below the 5-d Planck scale 
$M\equiv (16\pi G)^{-1/3}$ translates into $ML\gg 1$.
This is the regime where holographic duality is computationally effective.
In this regime the $\mbox{AdS}_5$ curvature is small compared with 
the 5-d Planck scale and one is thus 
justified in neglecting all higher-curvature terms in the 5-d gravitational 
action, and in equating the latter to the Einstein-Hilbert action with
cosmological constant. 
In physical examples, $G^{bare}_4 > 0$, thus $ML\gg 1$ is also sufficient 
to ensure that the momentum cutoff, $1/L$, is well below the 4-d Planck scale
$M_4\equiv (16\pi G_4)^{-1/2}$.
\item In eq.~(\ref{12}) we omitted terms that vanish with the
curvature faster than\\ $\int_M d^4 x \sqrt{-g^4}( R-2\lambda)$ or $W_{CFT}$.
\item If we denote by $K_{\mu\nu,\rho\sigma}$ the bare kinetic term of the 
4-d graviton, and if
we expand $\Gamma_H[g^4_{\mu\nu}]$ to quadratic order around a 
stationary point, obeying $\delta \Gamma_H /\delta g_{\mu\nu}^4=0$, 
we obtain the self-energy $\Sigma_{\mu\nu,\rho\sigma}$ as
\beq
{1\over 2}{\delta^2 \Gamma_H \over \delta g_{\mu\nu} \delta g_{\rho\sigma}}=
K_{\mu\nu,\rho\sigma} + \Sigma_{\mu\nu,\rho\sigma}.
\eeq{14}
\end{enumerate}

\subsection{Rigid and Gauged Holography II}
When the metric $g^4_{\mu\nu}$ can be expanded as 
\beq
g^4_{\mu\nu}= \exp(2\sigma)(\eta_{\mu\nu} + h_{\mu\nu}), 
\eeq{15}
i.e. when the 4-d background 
metric is conformally flat, we can show that flat-space holography implies 
curved-space holography in yet another way, particularly tailored to our 
background.
This new proof is given here not
only to convince the skeptic, but also to introduce some formulas that 
will be useful in the rest of this paper.

We assume that holography holds perturbatively around a 
flat background, i.e. that 
\beq
\Gamma_\epsilon^F[\eta_{\mu\nu} + h_{\mu\nu}] = 
W_{CFT}[\eta_{\mu\nu} + h_{\mu\nu}] +o(\epsilon),
\eeq{16} 
whenever both sides of this equation are expanded in powers of $h_{\mu\nu}$.

$W_{CFT}$ transforms as follows under the Weyl rescaling $g_{\mu\nu}=
\exp(2\omega) \bar{g}_{\mu\nu}$:
\beq
{\delta W_{CFT}\over \delta \omega}= c \{A[\bar{g}] -4
\bar{\rule{0cm}{0.25cm}\Box}_4 \omega\}, 
\qquad A[g]=
2R_{\mu\nu}R^{\mu\nu}- {2\over 3} R^2 -{2\over 3}\Box R, 
\qquad c=\,\mbox{constant}.
\eeq{17}
The computation of Henningson and Skenderis~\cite{hs} gives $c=L^3/128\pi G$ 
(see eqs.~(\ref{9},\ref{10})); $\Box_4$ is the operator~\cite{r}
\beq
\Box_4=\Box^2 + 2R^{\mu\nu}D_\mu D_\nu -{2\over 3} R\Box + {1\over 3} 
(\partial_\mu R)D^\mu.
\eeq{18}
Hereafter, an overbar will denote quantities computed with respect to the 
metric $\bar{g}_{\mu\nu}$.
$D_\mu$ is the covariant derivative and $\Box\equiv D_\mu D^\mu$. 
$\Box_4$ maps scalars of conformal 
weight zero into scalars of conformal weight four. 

As shown explicitly in ref.~\cite{hs}, ${\delta W_{CFT}/ \delta \omega}=
\delta \Gamma_\epsilon^F/ \delta \omega $. This equation, together with 
eq.~(\ref{16}) is sufficient to ensure that
\beq
\Gamma_\epsilon^F[\exp(2\sigma)(\eta_{\mu\nu} + h_{\mu\nu})]=
W_{CFT}[\exp(2\sigma)(\eta_{\mu\nu} + h_{\mu\nu})] +o(\epsilon).
\eeq{19}
This equation states that if holography is valid perturbatively around flat 
space, then it is also valid perturbatively around {\em any}
conformally flat background. 

Notice that the conformal transformation may map  
flat space into components joined only at their boundary. 
This phenomenon is manifest, for 
instance, when the flat metric $ds^2=-dt^2 +\sum_{i=1}^3dx^i dx^i$ 
is scaled by the Weyl transformation 
$\exp(\omega)=1/|x^3|$. The Weyl scaling given
here is induced by a 5-d diffeomorphism, as shown in appendix B. 
As we will see in section 4, in our background, 
the 4-d boundary of $\mbox{Ads}_5$ has two components, joined only 
at their edge.
This does not contradict the theorem of Witten and Yau~\cite{wy}, as 
the 4-d boundary has negative curvature. 
The very fact that the 4-d space we need, made of two $\mbox{AdS}_4$ 
components, can be connected to flat (Minkowsky) space by a Weyl 
transformation suggests that holography must hold also for our
background.  

The transformation property of the generating function,
given in eq.~(\ref{17}) gives us additional information on $W_{CFT}$ when
we expand $W_{CFT}$ to quadratic order in $h_{\mu\nu}$. 
By using known properties of $W_{CFT}$ (see for instance~\cite{ddi}) 
it is easy to see that
\beq
W_{CFT}[\eta_{\mu\nu} + h_{\mu\nu}]=
-{1\over 2}c\,\bar{C}_{\mu\nu}^{\rho\sigma}\log (\bar{\rule{0cm}{0.25cm}\Box}
/\mu^2)
\bar{C}^{\mu\nu}_{\rho\sigma} + O(h^3).
\eeq{20}
Here and below we omit the sign of integration in $d^4x$ in our formulae 
whenever unambiguous.
$C_{\mu\nu}^{\rho\sigma}$ is the Weyl tensor, and $\mu$ is a mass scale
introduced for dimensional reasons. Its arbitrariness reflects the fact that
$W_{CFT}$ is defined up to local Weyl-invariant terms.

We want to find the analog of eq.~(\ref{20}), but now when the metric is 
expanded to quadratic order in the fluctuations around an AdS background.
To do this we must promote $\log (\bar{\rule{0cm}{0.25cm}\Box}/\mu^2)$ 
to a linear operator
$F_{\mu\nu\,\alpha\beta}^{\rho\sigma\,\gamma\delta}(x,y)$, 
acting on tensors of same conformal weight and symmetries as
$C_{\mu\nu}^{\rho\sigma}$. $F$ must satisfy two properties
\bea
(\delta F/ \delta \omega)_{\mu\nu\,\alpha\beta}^{\rho\sigma\,\gamma\delta}
(x,y) &=& \delta^4(x,y) \delta_{\mu\nu}^{\gamma\delta} 
\delta_{\alpha\beta}^{\rho\sigma}, \label{21}\\ 
\lim_{g_{\mu\nu}\rightarrow \eta_{\mu\nu}} 
F_{\mu\nu\,\alpha\beta}^{\rho\sigma\,\gamma\delta} &=& -{1\over 2} 
\delta_{\mu\nu}^{\gamma\delta} \delta_{\alpha\beta}^{\rho\sigma}
\log(\bar{\rule{0cm}{0.25cm}\Box}/\mu^2).
\eea{22} 
The first equation generalizes the transformation property of 
$\log (\bar{\rule{0cm}{0.25cm}\Box}/\mu^2)$ under constant conformal transformations to 
arbitrary Weyl rescaling. The second equation is obviously necessary to
reproduce the expansion around flat space given in eq.~(\ref{20}).
That an $F$ obeying eqs.~(\ref{21},\ref{22}) exists has been argued 
in~\cite{d}, using the results in~\cite{brg}.
The form of $F$ is not uniquely fixed by the Weyl anomaly, since one can 
always add to $F$ Weyl-invariant terms. We will compute $F$ using 
the holographic correspondence in section 4.
Another possible $F$ is $-(1/4)\log \Delta/\mu^2$, 
where $\Delta$ is an operator
that respects the symmetry properties of $C_{\mu\nu}^{\rho\sigma}$, and
maps the Weyl tensor into a tensor of conformal weight 6~\cite{brg}. 

Now we are ready to compute 
$W_{CFT}[\exp(2\sigma)(\eta_{\mu\nu} + h_{\mu\nu})]$, expanded to quadratic 
order in $h$.
As before, we set $g_{\mu\nu}=\exp(2\sigma)(\eta_{\mu\nu}+h_{\mu\nu})$, 
$\bar{g}=\eta_{\mu\nu} +h_{\mu\nu}$.
We use eq.~(\ref{17}) to find
\beq
W_{CFT}[\bar{g}]= -c\{\sigma A[g] +2 \sigma \Box_4 \sigma \}+  
W_{CFT}[g].
\eeq{23}
Using the properties of $F$ given in eqs.~(\ref{21},\ref{22}), we also have
\beq
-{1\over 2}\bar{C} \log(\bar{\rule{0cm}{0.25cm}\Box}/\mu^2) \bar{C} = 
C F C - \sigma C^2 + O(h^3).
\eeq{24}
Combining eqs.~(\ref{20},\ref{23},\ref{24}) we find an expression for
$W_{CFT}[g]$:
\beq
W_{CFT}[g]=c\{CFC + \sigma(A[g]-C^2) + 2\sigma \Box_4 \sigma\} + O(h^3).
\eeq{25}
The explicit $\sigma$ dependence in this expression can be canceled by adding 
to it 
the trivially Weyl-invariant term $-(c/8)(\bar{A}-\bar{C}^2)
\bar{\rule{0cm}{0.25cm}\Box}_4^{-1} (\bar{A}-\bar{C}^2)$. We finally find
\beq
W_{CFT}[g]= c\left\{CFC -{1\over 8} (A-C^2)\Box_4^{-1}(A-C^2)\right\}+O(h^3).
\eeq{26}
This equation gives the generating functional of the 
conformal field theory, expanded to quadratic order 
around a conformally flat background. This equation also 
provides us with a useful decomposition of $W_{CFT}[g]$ into the ``universal''
Riegert~\cite{r} term, 
$-(c/8)(A-C^2)\Box_4^{-1}(A-C^2)$ obtained form integrating the 
conformal anomaly, and a model-dependent term, $c\,CFC$. 
In the next section we will show that 
the Riegert term {\em does not} give a mass to the the graviton.

\section{Analysis of the Riegert Term}
The upshot of this section is that the Riegert term does not give rise to a
mass term for the graviton. Readers not interested in the details of the 
computation can skip ahead to section 4, or to the end of this section for
an alternative proof of the statement.

The Riegert term can be rendered local by introducing an auxiliary field 
$\zeta$, in terms of which the term reads
\beq
W_R[g]\equiv -{c\over 8} (A[g]-C^2)\Box_4^{-1}(A[g]-C^2)= -{c\over 8}
\int d^4 x\sqrt{-g} [-\zeta \Box_4 \zeta + 2\zeta (A[g]-C^2)],
\eeq{27}  
where $\zeta$ obeys its own Euler-Lagrange equation: $\Box_4\zeta=A[g]-C^2$.
Notice that the combination $C^2-A$ is  $E_4+(2/3)\Box R$ and $E_4$ is
the Euler density 
$E_4=R_{\mu\nu\rho\sigma}R^{\mu\nu\rho\sigma} -4 R_{\mu\nu}R^{\mu\nu} + R^2$.

We want to show that the Riegert term does not give a mass to the graviton on
$\mbox{AdS}_4$.
To prove this statement, we will show that an Einstein space is always a
solution of 
the equations of motion that follow from the Einstein-Hilbert action modified
by the addition of the Riegert term: $S=S_{EH} + W_R$.
An Einstein space satisfies $R_{\mu\nu}=g_{\mu\nu}R/4$. When linearized around
an $\mbox{AdS}_4$ background, this equation describes the propagation of 
the standard massless graviton. 
Recall that in the KR model, there is {\em no} massless graviton excitation
propagating on $\mbox{AdS}_4$. This means that in the KR compactification, 
no Einstein space except $\mbox{AdS}_4$ can solve the equations of motion.

To compute the first variation of the action $S=S_{EH} + W_R$ 
around an Einstein space, we decompose the variation of the metric into
a trace part, $\phi$, a diffeomorphism part $\epsilon_\mu$, and a 
transverse-traceless part $\psi_{\mu\nu}$:
\beq
\delta g_{\mu\nu} = g_{\mu\nu} \phi + D_{(\mu} \epsilon_{\nu)} +
\psi_{\mu\nu}, \qquad D^\mu \psi_{\mu\nu}= \psi^\mu_\mu=0.
\eeq{28}

The variation of $S$ with respect to $\epsilon_\mu$ is the simplest to 
compute: by diffeomorphism invariance $\delta S / \delta \epsilon_\mu=0$ 
around any background.

The variation of $S$ with respect to $\phi$ is computed using eq.~(\ref{17})
and the definition of $W_R$
\beq
{\delta S \over \delta \phi} = {1\over 16\pi G_4} (R-4\lambda) 
+{c\over 2}(A-C^2). 
\eeq{29}
We can further simplify the calculation, now and later, by noticing that we
are looking for the quadratic part of the effective action around an 
AdS background. This means that we need to compute the variation of $S$ only 
to linear order in the fluctuation. Now, the Weyl tensor vanishes on 
AdS, since AdS is conformally flat. Moreover, by denoting with 
$\bar{R}$ the curvature of the AdS background, we find $A=-(1/6)\bar{R}^2 + 
O(h^2)$. Thus, 
\beq
{\delta S \over \delta \phi} = {1\over 16\pi G_4} (\bar{R}-4\lambda) 
-{c\over 12} \bar{R}^2 + O(h^2).
\eeq{30}
To the order in $h$ we are interested in, this equation just defines the
scalar curvature $\bar{R}$ in term of the ``bare'' cosmological constant 
$\lambda$~\footnote{The $\phi$ variation of the complete effective action, 
$\Gamma_H[g]$, which includes the term $CFC$, is 
$\delta \Gamma_H /\delta \phi = (1/ 16\pi G_4) (R-4\lambda) 
-(c/12)R^2$ on {\em any} Einstein metric. Eq.~(\ref{30}) is thus valid beyond
the quadratic approximation used in the text.}.

The variation of the Einstein-Hilber action with respect to $\psi_{\mu\nu}$ is
relatively simple.
From the variation of the Christoffel connection
\beq
\delta\Gamma^\rho_{\mu\nu}= {1\over 2}g^{\rho\lambda}(D_\mu 
\delta g_{\nu\lambda} + D_\nu \delta g_{\mu\lambda} -D_\lambda \delta 
g_{\mu\nu}), 
\eeq{31}
we have
\beq
\delta R= -\Box \delta g + D^\mu D^\nu \delta g_{\mu\nu} - \delta g_{\mu\nu} 
R^{\mu\nu}
\eeq{32}
Using the definition of $\psi_{\mu\nu}$ we find 
$\delta S_{EH}/\delta\psi_{\mu\nu}=0$ on an Einstein background (which obeys 
$R_{\mu\nu}=(1/4)g_{\mu\nu}R$).

The variation of $W_R$ is less straightforward.  

First of all, we must notice that the variation $\delta W_R /\delta \zeta$
vanishes, since $\zeta$ satisfies its own Euler-Lagrange equations. Thus, in 
eq.~(\ref{27}) we must only compute the explicit 
variation of the metric.

Let us compute now the variation of the term 
$2\int d^4x \sqrt{-g} \zeta (A[g]-C^2)$ in eq.~(\ref{27}).
Using
\beq
R^{\;\;\;\;\;\;\rho}_{\mu\nu\rho}= D_\mu \delta \Gamma^\rho_{\nu\lambda} - \mu
\leftrightarrow \nu,
\eeq{33}
we find, after a short calculation
\beq
\delta \int d^4x \sqrt{-g}\zeta (A[g]-C^2)= -4\int d^4 x \sqrt{-g}
D_\lambda D_\mu \zeta 
R^{\mu\nu\lambda}_{\;\;\;\;\;\;\rho} \psi_\mu^\rho 
\eeq{34}
up to terms that vanish on Einstein backgrounds.
For the purpose of our calculation, in this variation
we need to keep only terms linear in the fluctuation around the AdS background.
We use the $\mbox{AdS}_4$ metric 
$ds^2= dr^2 + \exp(2r/l)(-dt^2 + d\vec{x}^2)$ and the expansion~\cite{fps}
\beq
\dot{\zeta}\equiv {d \zeta \over d r} =-{4\over l} + O(h^2).
\eeq{35}
This definition implies a particular choice for the asymptotic behavior
of $\zeta$. That choice is an infrared regularization that we adopt as part
of the definition of the Riegert action.

Notice that the variation $\psi_{\mu\nu}$ is traceless with respect to the 
metric $\bar{g}_{\mu\nu} + h_{\mu\nu}$. Denoting by $\bar{\psi}_\mu^\nu$ etc.
quantities with indices raised and lowered with the background metric
$\bar{g}_{\mu\nu}$, after another short calculation, we arrive to
\beq
4 D_\lambda D_\mu \zeta R^{\mu\nu\lambda}_{\;\;\;\;\;\;\rho} \psi_\mu^\rho= 
{2\over l}\dot{\zeta} \bar{\triangle} h_{\mu\nu}
\bar{\psi}^{\mu\nu} +O(h^2)= -{8\over l^2} \bar{\triangle}h_{\mu\nu}
\bar{\psi}^{\mu\nu} +O(h^2).
\eeq{36}
Here, $\bar{\triangle}$ is the 3-d background Laplacian.

To compute the variation of the term $\int d^4 x \sqrt{-g}
\zeta \Box_4 \zeta$
we first integrate by part to find
\beq
\int d^4 x \sqrt{-g} \zeta \Box_4 \zeta = \int d^4 x \sqrt{-g}\left[
(\Box^2\zeta)^2 +{2\over 3} R \partial_\mu\zeta D^\mu\zeta -2 R^{\mu\nu}
\partial_\mu \zeta \partial_\nu \zeta \right].
\eeq{37}      
We further simplify the problem by the choice
$h_{3i}=h_{33}=0$ ($x^3\equiv r$, $i=0,1,2$). Likewise, we choose
$\psi_{33}=\psi_{3i}=0$. The rationale for this choice is that if a
mass term is generated by the Riegert action, it would make its variation
nonzero to linear order in $h$ even with respect to this restricted class
of variations.
This choice dramatically simplifies our calculations.
Indeed, the only term in eq.~(\ref{37}) that does not manifestly vanish is
\beq
-2\delta R^{33} 
\partial_3 \zeta \partial_3 \zeta= [\Box \psi^{33} -2D^3D_\mu\psi^{3\mu}
+2 R_{\lambda\;\;\rho}^{\;33\;} \psi^{\lambda \rho} 
-2 R_\rho^{(3}\psi^{3) \rho}] \dot{\zeta}^2. 
\eeq{38}
The computation of the various terms in the variation eq.~(\ref{38}) is
tedious but standard~\footnote{Most of the computations have been already done 
for the {\em de Sitter} background in~\cite{g}.}. 
On shell ($R_\mu^\nu=(1/4)\delta_\mu^\nu R$) we
have
\beq
R_\rho^{(3}\delta g^{3) \rho}= (1/4)R\psi_\mu^\mu=0.  
\eeq{39}
To linear order in the fluctuations around the AdS background, we find
\beq
R_{i\;\;j}^{\;33\;}\psi^{ij}= \left( -{1\over 2} 
\stackrel{..}{\bar{h}}_i^l 
\bar{g}_{lj} - {1\over l} \dot{\bar{h}}_i^l\bar{g}_{lj}\right)\bar{\psi}^{ij} 
+O(h^2).
\eeq{40}
As earlier, an overbar indicates that indices are raised and lowered with 
the background metric $\bar{g}_{\mu\nu}$.
Also, in the same approximation,
\bea
\Box \psi^{33} &=& {2\over l} \dot{\bar{h}}_i^j \bar{\psi}_j^i,\label{41}
\\
2D^3D_\mu \psi^{3\mu} &=& -\partial_r(\dot{\bar{h}}^i_j\bar{\psi}^i_i).
\eea{41a}
This is the last result we need, since by combining 
eqs.~(\ref{39},\ref{40},\ref{41},\ref{41a}) with eq.~(\ref{36}) 
and the definition of $W_R$ in eq.~(\ref{27}) we conclude
\beq
\delta W_R =-{c\over l^2}\int d^4x \sqrt{-\bar{g}}
\left(\stackrel{..}{\bar{h}}_i^l + {3\over l}\dot{\bar{h}}_i^j + 
\bar{\triangle}\bar{h}^i_j\right)\bar{\psi}^j_i  + O(h^2).
\eeq{42}
In this equation we have integrated by part the term $2D^3D_\mu \psi^{3\mu}$.
Eq~(\ref{42}) is just the standard equation of motion of the massless 
graviton in AdS; therefore, the mass term can not be generated  by $W_R$.

An alternative proof of this statement can be obtained by using the results of
Mazur and Mottola~\cite{mm}.
In that paper, Mazur and Mottola find 
an expression for the variation of the Riegert action 
by extending the computation of the conformal anomaly to $4+\epsilon$ 
dimensions, and taking the limit $\epsilon\rightarrow 0$. The expression, given
in their eq.~(5.12), contains an auxiliary tensor, called by them 
$C_{\mu\nu}$. Ref.~\cite{mm} contains in its appendix C 
an explicit formula for $C_{\mu\nu}$. Using that formula, and recalling that 
in our background the square of the Weyl tensor vanishes up to quadratic 
order in the fluctuation $h_{\mu\nu}$, we arrive again at eq.~(\ref{42}).  

\section{The Holographic Computation}
Since the mass term found in~\cite{kr} cannot come from the Riegert term, we 
have to look for it in the {\em model-dependent} term of the action, whose
quadratic part is proportional to $CFC$. To compute the operator 
$F_{\mu\nu\,\alpha\beta}^{\rho\sigma\,\gamma\delta}(x,y)$ we use the 
``gauged'' holographic duality introduced in section 2.

The generating functional $W_{CFT}$ computed holographically 
automatically obeys all properties expected in field theory, in particular it
obeys the Ward identities of conformal symmetry. The reason behind this is that
conformal transformations in the 4-d boundary theory are just special general
coordinate transformations of the 5-d theory, that obeys general covariance 
(see for instance~\cite{isty} and appendix A).

In the rest of this section, we will study only the transverse-traceless part
of the metric fluctuation. To determine the true value of the mass term, as 
opposed to effects due to a trivial renormalization of the cosmological 
constant we must treat more precisely the Gibbons-Hawking term.

\subsection{The Gibbons-Hawking Term and the Effective Action $\Gamma_H$}
Let us write the 5-d metric $g_{mn}$ ($m,n=0,1,2,3,4$) in eq.~(\ref{2}) as
\beq
ds^2= \exp(2A)[dz^2 + (\bar{g}_{\mu\nu} + h_{\mu\nu})dx^\mu dx^\nu],\qquad
A=-\log \cos z, \qquad \mu,\nu=0,1,2,3.
\eeq{43}
In this section we set $L=1$.
In this metric, the Gibbons-Hawking term, given in eq.~(\ref{7}), 
can be written as
\beq
S_{GH}={1\over 8\pi G}\int d^4 x \sqrt{-\det(\bar{g} +h)}\exp(4A)
{1\over \sqrt{-g}} \partial_z (\sqrt{-g} n^z), \qquad n^z=\exp(-A).
\eeq{44}
Expanding this equation to quadratic order in $h$, after a short computation
we arrive at~\footnote{The computation is standard and can be found, for 
instance in~\cite{hhr}.}
\beq
S_{GH}=-{1\over 8\pi G}\int d^4 x \sqrt{-\bar{g}}\exp(3A)
\left[ {1\over 2} h_\mu^\nu h_\nu^\mu \dot{A}
 + \dot{h}_\mu^\nu h_\nu^\mu\right], \qquad \cdot \equiv \partial/\partial z.
\eeq{45}
Here and elsewhere in this section indices are raised and lowered with the
4-d metric $\bar{g}_{\mu\nu}$.
The quadratic part of the Einstein-Hilbert action, computed  on-shell, 
reduces to a boundary term
\beq
S_{EH}= {1\over 16\pi G}\int d^4 x \sqrt{-\bar{g}}
\exp(3A)\left[ {3\over 4} \dot{h}h 
+{1\over 2} h^2 \dot{A}\right], \qquad 
h^2\equiv h_\mu^\nu h_\nu^\mu \;\mbox{etc}.
\eeq{46}     
We put the boundary at $\cos z=\varepsilon$.

Here, it is convenient to keep $h_{\mu\nu}$ as our 4-d metric instead of 
performing the rescaling of subsection 2.2.
The sum of the Einstein-Hilbert action and the Gibbons-Hawking term gives
the regularized action $\Gamma_\varepsilon$. Expanded to
quadratic order in $h_{\mu\nu}$ it reads
\beq
\Gamma_\varepsilon=-\left.{1\over 16\pi G}\int d^4 x \sqrt{-\bar{g}}\exp(3A)
\left( {1\over 4} \dot{h}h +{3\over 2} \dot{A} h^2\right)
\right|_{\cos z=\varepsilon}.
\eeq{47}
Before jumping to the conclusion that $(3/2) \dot{A} h^2$ is a mass term
we must recall that 4-d counterterms can be added to 
$\Gamma_\varepsilon$. As we mentioned earlier, these counterterms cancel the 
$\Gamma_\varepsilon$ divergences in the limit $\varepsilon\rightarrow 0$. Here,
$\varepsilon$ is kept finite, but the counterterms can still be added. 
The mass term is an artifact if it can be canceled by 
appropriately choosing them.

To properly identifying the meaning of the term $(3/2) \dot{A} h^2$, we
notice that 
\beq
\dot{A}=\tan z ={1\over \cos z} -{1\over 2} \cos z -{1\over 8} (\cos z)^3 +
O[(\cos z)^5].
\eeq{48}
Using this expansion we can write eq.~(\ref{47}) as
\beq
\Gamma_\varepsilon={1\over 16\pi G}\int d^4 x \sqrt{-\bar{g}}
\left[ -{1\over 4\varepsilon^3 } \dot{h}h +\left(-{3\over 2\varepsilon^4} +
{3\over 4 \varepsilon^2} +{3\over 16} \right)h^2\right] +O(\varepsilon^2).
\eeq{49}
It is now easy to identify the various terms in this equation. Terms that
diverge in the $\varepsilon\rightarrow 0$ limit are local; they do not
give a mass term, instead, they renormalize
the cosmological constant and Einstein term in the 4-d action. In particular,
the term $-(3/2)\varepsilon^{-4}h^2$ renormalizes the cosmological constant
as 
$\delta [(16\pi G_4)^{-1}\lambda]=-3(16\pi G)^{-1}\varepsilon^{-4}$. 
The term $(3/4)\varepsilon^{-2}h^2$ renormalizes the Newton constant
as $\delta [(16\pi G_4)^{-1}]= (32\pi G)^{-1}\varepsilon^{-2}$.

The finite term $(3/16)h^2$ is not necessarily a true mass term, since it 
may arise from a finite renormalization of the 4-d cosmological constant. 
To find if this is the case, let us compute the variation of the holographic
effective action $\Gamma_H$ under a scale transformation. Using the definition 
in eq.~(\ref{12}) and eq.~(\ref{17}) we find
\beq
{1\over \sqrt{-g}}g_{\mu\nu}{\delta \Gamma_H \over \delta g_{\mu\nu}}=
{1\over 16\pi G_4} (-4\lambda + R)  -{c\over 12} R^2,
\qquad c={1\over 128\pi G}.      
\eeq{50}
The background metric $\bar{g}_{\mu\nu}$ solves by construction the 
equations of motion
$\delta \Gamma_H / \delta g_{\mu\nu}=0$. Our metric $\bar{g}_{\mu\nu}$ 
has scalar curvature $\bar{R}=-12$. Recall from the discussion at the end of 
subsection 2.2 that to guarantee the validity of the 
holographic computation, both the 4-d and 5-d curvatures must be smaller than 
their Newton constants. In our units this means $|c\bar{R}^2/16| \ll
|\bar{R}/16\pi G_4|\ll 1$. It means also that we can expand the cosmological 
constant $\lambda$ as $\lambda=\bar{\lambda} + \delta \lambda$;
$\bar{\lambda}=-3$, $\delta\lambda\ll 1$.
Substituting in eq.~(\ref{50}) we find
\beq
{1\over 16\pi G_4}\delta \lambda = -{1\over 16 \pi G} {3\over 8}. 
\eeq{51}
This finite renormalization of $\lambda$ generates an apparent mass term equal
to 
\beq
{1\over 32\pi G_4}\delta \lambda h^2 = -{1\over 16 \pi G} {3\over 16}h^2.
\eeq{52}
This is exactly the term necessary to cancel the $\varepsilon$-independent
term in eq.~(\ref{49}).

We have concluded at last that $\Gamma_H$ expanded to quadratic order in
the transverse-traceless fluctuation $h_{\mu\nu}$ is: 
\beq
\Gamma_H  = {1\over 16\pi G_4} \int d^4x \sqrt{-\bar{g}} {1 \over 4}\left[ 
h (-\triangle_L^{(2)} + 2\bar{\lambda}) h\right]
- {1\over 16\pi G}\int d^4 x \sqrt{-\bar{g}}{1\over 4\varepsilon^3 } \dot{h}h 
\eeq{53}
Here, $\triangle_L^{(2)}$ is the Lichnerowicz operator on symmetric 
tensors~\cite{l}. On our AdS background, and on transverse-traceless tensors  
it reads
\beq
\triangle_L^{(2)}h_{\mu\nu}= -\Box h_{\mu\nu} +{8\bar{\lambda}\over 3}
h_{\mu\nu}.
\eeq{54}

We can perform an easy check on our result: substituting into it the
non-normalizable graviton zero mode, eq.~(\ref{53}) must give
the action of a massless 
spin 2. This is obviously correct as the non-normalizable zero mode 
is independent of $z$, and thus it obeys $\dot{h}_{\mu\nu}=0$.

\subsection{The Dressed Graviton Propagator}
We are at last ready to compute the self-energy 
$\Sigma_{\mu\nu,\rho\sigma}$ introduced in eq.~(\ref{14}). We use the 
holographic duality so that the self-energy we are looking for is given by
\beq
h^{\mu\nu}\Sigma_{\mu\nu,\rho\sigma}h^{\rho\sigma}= 
- {1\over 16\pi G}\int d^4 x \sqrt{-\bar{g}}{1\over 4\varepsilon^3 } \dot{h}h.
\eeq{55} 
The equation obeyed by $h_{\mu\nu}$ is~\cite{kr,gs}
\beq
\left[-\partial_z(\cos z)^{-3}\partial_z - (\cos z)^{-3} \left(\Box 
-{2\bar{\lambda}\over 3}\right)\right]h_{\mu\nu}=0.
\eeq{56}
Let us decompose $h_{\mu\nu}$ into eigenstates of $\Box$, denoted by
$h^m_{\mu\nu}(x)$:
\beq
h_{\mu\nu}(x,z)=\sum_m h^m_{\mu\nu}(x)H^m(z), \qquad 
[\Box -(2\bar{\lambda}/3)]h^m_{\mu\nu}(x)=m^2 h^m_{\mu\nu}(x).
\eeq{57}   

The differential equation for $H^m(z)$ can be transformed into a 
 standard hypergeometric form by the change of variable $y=(\cos z)^2$.
In terms of the new variable, the equation reads 
\beq
\left[(1-y)y\partial_y^2 + \left(-1+{1\over2}y\right)\partial_y +{m^2\over4} 
\right]H^m(z)=0.
\eeq{58}   
Its two independent solutions are 
\bea
H_1^m &=& [\psi(1) +\psi(3) -\psi(a+2) -\psi(b+2)]
y^2 F(a+2, b+2;3;y), \label{59}\\
H_2^m &=& y^2 F(a+2,b+2;3;y) - {2\over ab(a+1)(b+1)} +{2\over (a+1)(b+1)}y
+ \nonumber \\ && 
+ y^2 \sum_{n=1}^\infty y^n{\Gamma(a+2+n)\Gamma(b+2+n)\Gamma(3)\over
\Gamma(a+2)\Gamma(b+2)\Gamma(3+n)\Gamma(n+1)}[\psi(a+2+n) -\psi(a+2) + 
\nonumber \\ &&
+\psi(b+2+n) -\psi(b+2) -\psi(3+n) + \psi(3) -\psi(n+1) +\psi(1)], \label{60}
\\ && ab=-m^2/4,\qquad a+b +1 =-1/2.
\eea{61}
See~\cite{gr} for notations.  
Since our change of variables $y=(\cos z)^2$ is 2-to-1, 
the interval $0\geq y \geq 1$ actually covers two distinct domains,
$[-\pi/2,0]$ and $[0,\pi/2]$. The solution we are looking for has two different
expansions in the two domains; namely
\bea
H^m &=& \alpha H_1^m + \beta H_2^m
\qquad \mbox{in } [-\pi/2,0], \label{62} \\
H^m &=& \alpha' H_1^m + \beta' H_2^m \qquad \mbox{in } [0,\pi/2].
\eea{63}
For the holographic computation we impose the following boundary conditions.
\begin{enumerate}
\item
At ``our end'' of the $\mbox{AdS}_5$ space, $z=\pi/2$ ($y=0$) we set 
$H^m(\pi/2)=1$~\footnote{We have been sloppy here, as the
proper normalization condition is $H^m=1$ at $\cos z=\varepsilon$. 
The proper normalization condition can be obtained from that used in the
text by rescaling $H^m$ as follows $H^m(y)\rightarrow 
H^m(y)/H^m(\sqrt{\varepsilon})$. This rescaling does not significantly affect
our computations and results.}.
By this choice, the boundary value of the fluctuation 
$h_{\mu\nu}(x,z)$ becomes a sum over 4-d free fields of AdS mass $m$
\beq
h_{\mu\nu}(x,\pi/2)=\sum_m h^m_{\mu\nu}(x).
\eeq{64}   
\item At the ``other end'' of $\mbox{AdS}_5$, $z=-\pi/2$ (again $y=0$)
we set $H^m(-\pi/2)=0$. With this boundary condition no field can leak
out of the $\mbox{AdS}_5$ space. This boundary condition also ensures that
the $H^m(z)$ are normalizable, and {\em it removes from the spectrum} 
the zero mode $H^0(z)=1$. This is the crucial choice that gives rise to a
graviton mass. 
\item At $z=0$ ($y=1$) we have to match the two expansions for $H^m(z)$.
This is done by matching $H^m(z)$ and its first derivative $dH^m(z)/dz$.
Notice that in the interval $[-\pi/2,0]$ we have
\beq
\partial_z = 2\sqrt{y(1-y)}\partial_y,
\eeq{65}
while in $[0,\pi/2]$ we have
\beq
\partial_z = -2\sqrt{y(1-y)}\partial_y.
\eeq{66}
This means that the matching conditions at $y=1$ are
\bea
&& \alpha H^m_1(1) + \beta H^m_2(1) = \alpha' H^m_1(1) + \beta' H^m_2(1),
\label{67} \\ &&
\lim_{y\rightarrow 1} \sqrt{(1-y)}\left(\alpha {d H^m_1 \over dy} + 
\beta {d H^m_2 \over dy}+\alpha' {d H^m_1 \over dy}  + 
\beta'{d H^m_2 \over dy}\right)_{y=1}=0.
\eea{68}
\end{enumerate}
The boundary condition at $z=\pi/2$ sets
\beq
\beta'= -{ab(a+1)(b+1)\over 2}.
\eeq{69}
The boundary condition at $z=-\pi/2$ implies
\beq
\beta=0.
\eeq{70}

To analyze the matching conditions at $y=1$ we use standard identities
among hypergeometric functions to find
\bea
H_1^m(y)&=& A +\sqrt{1-y}B + O(1-y), \qquad 
H_2^m(y)= C +\sqrt{1-y}D + O(1-y), \label{71} \\
A&=& [\psi(1)+\psi(3)-\psi(a+2)-\psi(b+2)]{\Gamma(3)\Gamma(1/2)\over
\Gamma(1-a)\Gamma(1-b)}, \label{71a} \\ 
C&=&-2F(a,b;a+b+2;1){\Gamma(a)\Gamma(b)\over
\Gamma(1/2)}, \label{72}\\
B&=&D=[\psi(1)+\psi(3)-\psi(a+2)-\psi(b+2)]{\Gamma(3)\Gamma(-1/2)\over
\Gamma(2+a)\Gamma(2+b)}.
\eea{73}
Since $B=D$, matching the $z$-derivative gives 
$\alpha+\alpha'+\beta'=0$. Finally, eq.~(\ref{67}) gives
\beq
\alpha'= {ab(a+1)(b+1)\over 4}\left( 1 + {C\over A}\right).
\eeq{74}.

After applying some further hypergeometric identities we find that 
$H^m(z)$ near $z=\pi/2$ can be written as  
\bea
H^m(y)&=& 1 -aby -{ab(a+1)(b+1)\over 2}y^2\log (y/\sqrt{e}) +
y^2 {\cal F}(m^2) +O(y^3), \label{74a} \\
{\cal F}(m^2) &=& {ab(a+1)(b+1)\over 4} \Bigg\{ 
[\psi(1)+\psi(3)-\psi(a+2)-\psi(b+2)] +
 \nonumber \\ && 
-{\Gamma(a)\Gamma(b)\Gamma(1-a)\Gamma(1-b)\over 
\Gamma(1/2-a)\Gamma(1/2-b)\Gamma(3)\Gamma(1/2)} \Bigg\}.
\eea{75}

Three important checks can be performed on this formula.
\begin{enumerate}
\item 
In the limit $m^2\rightarrow 0$, $a\rightarrow 0$ and $b\rightarrow -3/2$
(see the definition of $a,b$ in eq.~(\ref{61})). The formula for
$H^0(y)$ simplifies dramatically: 
\beq
H^0(y)= 1-{3\over 16}y^2+O(y^3).
\eeq{76}
This equation must be compared with the explicit, elementary solution of 
eq.~(\ref{56}) for $m^2=0$
\beq
H^0(z)= {1\over 2} +{3\over 4} \sin z -{1\over 4} (\sin z)^3= 
{1\over 2} +{3\over 4} (1-y)^{1/2} -{1\over 4} (1-y)^{3/2}=
1-{3\over 16}y^2+O(y^3).
\eeq{77}

\item
In the far Euclidean region eq.~(\ref{75}) must reproduce the flat space 
result.
Setting $-m^2=p^2$ we have $b^*=a=-(3/4) +ip +O(p^{-1})$. 
Using the asymptotic formula
\beq
\left|{\Gamma[-(3/4) +ip]\Gamma[(7/4) +ip]\over \Gamma[(5/4) +ip]}\right|^2
= 2\pi \exp(-\pi |p|)[|p|^{1/2} + O(|p|^{-1/2})],
\eeq{78}
we find 
\bea
H^{ip}(y) &=& 1 + {p^2\over 4}y  - 
{p^2\over 8}\left({p^2\over 4}-{1\over 2}\right)y^2\log (y/\sqrt{e})+
\nonumber \\ && 
-\left[{p^2\over 8}\left({p^2\over 4}-{1\over 2}\right)\log |p| +P(p^2)\right]
y^2 +O(y^3).
\eea{79}
Here $P(p^2)$ is a polynomial in $p^2$.
For $p^2 \gg 1$ this formula does indeed
coincide with its flat space analog (see~\cite{gkp}).

\item
The pure $AdS_4$ mass spectrum is recovered when we ask that $H^m$ vanishes
also at $z=\pi/2$. This give the conditions
\beq
\hat{A}\equiv 
A/[\psi(1)+\psi(3)-\psi(a+2)-\psi(b+2)]={\Gamma(3)\Gamma(1/2)\over
\Gamma(1-a)\Gamma(1-b)}=0,
\eeq{80a}
or 
\beq 
\hat{B}\equiv 
B/[\psi(1)+\psi(3)-\psi(a+2)-\psi(b+2)]={\Gamma(3)\Gamma(-1/2)\over
\Gamma(2+a)\Gamma(2+b)}=0. 
\eeq{80}
A zeros of either $\hat{A}$ or $\hat{B}$ arises only when the denominator in
either eq.~(\ref{80a}) or eq.~(\ref{80}) has a pole. This happens for 
$2+b=0,-1,-2,..$ and $1-a=0,-1,-2,..$. Recalling the definition of
$a $ and $b$ we arrive at the equation
\beq
m^2=n(n+3), \qquad n=1,2,3,...
\eeq{81}
These are precisely the masses of the 4-d spin-2 excitations that
make up the spectrum of the 5-d massless 
graviton~\cite{kr}.~\footnote{A pellucid introduction to  mass spectra in
$\mbox{AdS}_4$ can be found 
in~\cite{n}; \cite{ffz}  contains 
an up to date discussion of harmonic analysis for AdS spaces in
various dimensions.}
\end{enumerate}

Now we are ready to compute the self energy defined in eq.~(\ref{55}).  
We use the expansion eq.~(\ref{57}) and the formula for $H^m(z)$ given in
eqs.~(\ref{74a},\ref{75}). After using 
\beq
{\partial H^m \over \partial z}= 
-4y^{3/2}{\cal F}(m^2) + P(m^2) +O(y^{5/2}),
\qquad P(m^2)=\,
\mbox{polynomial in}\; m^2,
\eeq{82}
we find
\beq
h^{m\,\mu\nu}\Sigma_{\mu\nu,\rho\sigma}h^{m\,\rho\sigma}= 
 {1\over 16\pi G}\int d^4 x \sqrt{-\bar{g}} h_{\mu\nu}^m(x) 
{\cal F}(m^2)
h^{m\,\mu\nu}(x) + Q(m^2) + O(\varepsilon^2).
\eeq{83}
$Q(m^2)$ is another polynomial in $m^2$. It is not difficult to show that it
vanishes at $m^2=0$, so that it effects only a renormalization of
the Newton constant and the coupling constants for higher-derivative
local terms, irrelevant at low energies.

Now we come to the central result of this paper. As predicted by Karch and 
Randall, we find that: 
\vskip .1in
{\em The 4-d graviton acquires a nonzero mass $O(\lambda^2)$.}
\vskip .1in
To prove this we recall the definition of the effective action $\Gamma_H$, 
eq.~(\ref{53}).  On the transverse-traceless field $h^m_{\mu\nu}$ it reads
\beq
\Gamma_H  = {1\over 16\pi G_4} \int d^4x \sqrt{-\bar{g}} {1 \over 4}
h^{m\,\nu}_\mu m^2 h^{m\,\mu}_\nu
+{1\over 16\pi G}\int d^4 x \sqrt{-\bar{g}}h^{m\,\nu}_\mu 
 {\cal F}(m^2)  h^{m\,\mu}_\nu.
\eeq{84}

For small $m^2$, ${\cal F}(m^2)\approx -(3/16)$, so that the
pole of the propagator is shifted to a nonzero (positive) value.
In other words, the graviton gets a mass $m^2= (3G_4/4G)$!

Let us conclude this section with a few comments
\begin{enumerate}
\item
Recall that in this section we measured $G$ in units such that $L=1$, 
while $G_4$, the 4-d Newton constant, was measured in units such that $l=1$. 
By re-introducing the 5-d AdS radius, the 5-d Newton constant scales 
as $ G\rightarrow G/L^3$. Likewise, by re-introducing the 
the 4-d radius, the 4-d Newton constant scales as $G_4\rightarrow G_4/l^2$.
After these rescalings, recalling that the 4-d cosmological constant is
$\lambda=-3/l^2$, we find that the graviton mass assumes the 
more familiar form
\beq
m^2 = {1\over 12}{G_4 L^3\over G}\lambda^2.
\eeq{87}  

Notice that when the Newton constant is completely induced by the CFT 
($G^{bare}_4=\infty $ in eq.~(\ref{13})) we find
$m^2= L^2\lambda^2/6$.

\item
With our holographic calculation, we have found an explicit formula for the
term $CFC$: thanks to the asymptotic expansion eq.~(\ref{79}) we see that
${\cal F}(m^2)$ has all the right properties of the operator $F$; namely, it
obeys eqs.~(\ref{21},\ref{22}).
Of course, ${\cal F}(m^2)$ is a gauge-fixed version of $F$. In fact, the
boundary condition $H^m(-\pi/2)=0$, together with the metric choice
in eq.~(\ref{43}), $g_{\mu4}=0$, $g_{44}=\exp(2A)$, 
completely fixes the gauge for $h_{\mu\nu}$, as no diffeomorphism leaves
both the metric and the boundary condition invariant. 
Invariance  under diffeomorphisms and Weyl transformations of the 
non-gauge-fixed version of $\Gamma_H$ is nevertheless guaranteed because they 
both come from 5-d diffeomorphisms~\cite{isty}.
A more complete treatment of gauge fixing and the counting of degrees of 
freedom is given in appendix A. 

\item 
Eq.~(\ref{84}) gives not only the almost-massless graviton but also the tower
of massive Kaluza-Klein states found in~\cite{kr}. The easiest way to see this
is to expand ${\cal F}(m^2)$ 
near one of its massive poles as
\beq
{\cal F}(m^2) \approx {F_i \over m^2-m_i^2}.
\eeq{87a}
the rescaling introduced above tells us that $m_i^2$  and
$F_i$ are $O(\lambda)$. To find the Kaluza-Klein states we must set
\beq
{1\over 16\pi G_4} {m^2 l^4\over 4} + 
{L^3\over 16\pi G} {F_i \over m^2-m_i^2}=0.
\eeq{87b} 
By writing $m^2=m_i^2 + \delta m^2$, $\delta m^2 \ll m_i^2$ we solve
eq.~(\ref{87b}) as
\beq
\delta m^2\approx  - {4G_4 L^3 F_i \over l^4 G m_i^2}.
\eeq{87c}
To ensure the consistency of the approximation used here we must have
$G_4 \lambda \ll GL^{-3}$. In physical models where $G_4^{bare}>0$,
$G_4 < 2GL^{-1}$ so that the consistency condition is always satisfied,
since in that case it becomes $\lambda \ll L^{-2}$, and the 
4-d cosmological constant is always much smaller than the 5-d one.
\end{enumerate}
\section{Coda: Other Models, Other Spectra}
Variations on the KR compactification have been considered in the literature.
In particular, \cite{kmp2} considers the AdS equivalent of RSI, in which 
another (positive-tension) brane is set at $z=-\pi (/2)L+ \epsilon'$. 
In that case 
the graviton zero mode is normalizable, so that the spectrum contains {\em two}
spin-2 states much lighter than $\lambda$: one massless, the other with mass
$O(L^2\lambda^2)$. We propose to interpret the ``far'' brane as an effective
description of an infrared cutoff, as in the holographic interpretation
of the RSI model~\cite{ahpr}. While there are some aspects of this 
identification that are somewhat puzzling --e.g. what does it mean that the
far brane has positive tension?-- 
one by-product of this identification is satisfying.
Namely, with an infrared cutoff $\mu$, 
the generating functional of any 4-d field 
theory can always be expanded at low energies in terms of
local functions of the 4-d metric:
\beq
W[g]= \int d^4 x \sqrt{g} \sum_{n=0}^\infty \mu^{4-2n} O^{(2n)}(g),
\eeq{88} 
where $O^{(2n)}$ denotes local operators of dimension $2n$. This expansion 
guarantees that in an AdS background there always exists a 
massless graviton~\cite{kkr}. 

To see this, we notice that the effective action  
\beq
\Gamma[g]= {1\over 16\pi G_4}
\int_M d^4 x \sqrt{-g}( R-2\lambda) + W[g], 
\eeq{89}
is built with polynomials in the scalar curvature, $R$, the tensor
$R_{\mu\nu}-g_{\mu\nu}R/4$, and the Weyl tensor $C^{\mu\nu}_{\rho\sigma}$.
Expanding to quadratic order around an AdS solution of the equations of
motion of $\Gamma$, $\delta \Gamma/ \delta g_{\mu\nu}=0$, one finds 
\beq
\Gamma[g]={1\over 16\pi G_4}
\int_M d^4 x \sqrt{-\bar{g}} h^{\mu\nu} L(\triangle_L^{(2)}) 
(\triangle_L^{(2)}-\bar{R}/2)h_{\mu\nu} +O(h^3).
\eeq{90}
As before, the AdS background is $\bar{g}_{\mu\nu}$ and the fluctuation is
$h_{\mu\nu}$; $ L(\triangle_L^{(2)})$ 
is a polynomial in $\triangle_L^{(2)} $. 
Eq.~(\ref{90}) makes it manifest that the massless
graviton --obeying $(\triangle_L^{(2)}-\bar{R}/2)h_{\mu\nu}=0$-- 
still solves the linearized equations of motion.

Alternatively, the far brane could be interpreted as a CFT on another 
$\mbox{AdS}_4$ space, joined with ``our'' $\mbox{AdS}_4$ at its 
boundary, $S_2\times R$. In this case, the challenge is to understand whether 
the presence of two light gravitons can be seen as due to a peculiar 
boundary interaction between the two universes. 
The very possibility of this effect is probably due to the fact,
peculiar to AdS spaces, that null rays take a finite coordinate time to 
complete the round trip from the interior to the boundary and back.

At this point we need to repeat that in this paper we have argued that
holography works even when the boundary is made of disconnected components,
provided that we give appropriate boundary conditions on the metric.
If this is the case, the peculiar phenomena described in this paper (i.e.
the massive graviton) should be interpretable, as suggested 
in the previous paragraph, as due to a non-standard behavior of the 4-d fields
at the boundary of the 4-d space.

Another interesting question is whether the function ${\cal F}$ we found
in section 4 is generic in CFTs. In particular, is the graviton mass a
universal feature of CFTs coupled to gravity in AdS or 
is it an accident of our holographic computation?

To answer this question, it would be interesting to compute ${\cal F}$ in
a theory as far removed as possible from the strongly-coupled CFT studied here
using its holographic dual. For instance, a conformally-coupled 
free scalar could be an excellent test-ground. 
The free scalar computation would also clarify the effect of the
$\mbox{AdS}_4$ boundary conditions on the mass spectrum.
Anyway, even before any computation, it
is easy to see that we cannot rule out the possibility that the mass term is
peculiar to holographic models. Indeed, one can exhibit other operators
that obey eqs.~(\ref{21},\ref{22}) besides ${\cal F}$. One such example was
mentioned in section 2: $F=-(1/4)\log \Delta $. The operator $\Delta$ is the
conformally covariant completion of $\Box^2$ that maps conformal 
tensors with the symmetries of the Weyl tensor into conformal
tensors with the same symmetries and weight 6~\cite{d,brg}.
\vskip .2in  
{\bf Note Added in Proof}\vskip .1in
\noindent
After this paper was accepted for publication, it was pointed out to us that 
in 3 dimensions a {\em local} modification of the 
Einstein-Hilbert action exists, that gives a nonzero mass to the graviton 
while preserving general covariance~\cite{dj}. The mechanism of ref.~\cite{dj}
is peculiar to 3 dimensions. 
\vskip .2in
{\bf Acknowledgments}\vskip .1in
\noindent
We would like to thank LBL, where part of this research has been done, 
for its hospitality and support; L. Randall for many discussions on KR, 
related and unrelated topics and S. Deser for, among other things, 
bringing to light ref.~\cite{brg}.  
This work is supported in part by NSF grant PHY-0070787.

\section*{Appendix A: Diffeomorphisms and Gauge Fixing}
\renewcommand{\theequation}{A.\arabic{equation}}
\setcounter{equation}{0}
In the holographic setting,
4-d diffeomorphisms and conformal transformations both come from 5-d 
diffeomorphisms that keep the 5-d metric $g_{mn}$ in the gauge
\beq
g_{44}=\exp(2A), \qquad g_{\mu4}=0, \qquad \mu,\nu=0,..,3.
\eeq{a2}
The definition of the 5-d metric is
\beq
ds^2 = \exp[2A(z)] g_{mn}dx^m dx^n, \qquad m,n=0,..,4,
\qquad x^4\equiv z.
\eeq{a1}
In the KR model $A(z)=-\cos(z/L)$, in RSII $A(z)=-\log(z/L)$.
The gauge-preserving diffeomorphisms act on the 4-d metric as
\beq
\delta g_{\mu\nu}(x,z)= D_\mu \zeta_\nu(x,z) + D_\nu \zeta_\mu (x,z)
+ 2\dot{A} g_{\mu\nu}\zeta_5(x,z).
\eeq{a3}
The gauge choice eq.~(\ref{a2}) gives (see e.g.~\cite{isty,kr,kkr})
\bea
\zeta_5 &=&  \omega(x),\label{a4} \\ \zeta_\mu &=& G(z) D_\mu \omega(x)
+ \epsilon_\mu(x), \qquad G=\int dz \exp(-2A) ,
\eea{a5}
where both $\omega(x)$ and $\epsilon_\mu(x)$ are independent of $z$.
From its action on $g_{\mu\nu}$, it is clear that $\omega$ is a 4-d Weyl 
transformation and $\epsilon_\mu$ a 4-d diffeomorphism.
As explained in~\cite{kr,kkr} the general solution of the equations of motion
for the metric $g_{\mu\nu}=\exp(2A)(\bar{g}_{\mu\nu} + h_{\mu\nu})$ is, to
linear order in $h_{\mu\nu}$,
\bea
h_{\mu\nu}(x,z)&=& h_{\mu\nu}^{TT}(x,z) + 2GD_\mu D_\nu\Phi(x)  + 
2\dot{A} \bar{g}_{\mu\nu}\Phi(x), \label{a6}
\\
\left( \Box + {4\over 3}\lambda\right)\Phi &=& 0 
\eea{a7}
The field $\Phi(x)$, independent of $z$, can be canceled by setting 
$\omega=-\Phi$~\cite{kr}.

The transverse-traceless field  $h_{\mu\nu}^{TT}(x,z)$ can be further 
decomposed as
\bea
h_{\mu\nu}^{TT}(x,z) &=& \sum_m h^m_{\mu\nu}(x)H^m(z) + D_\mu A_\nu(x)
+ D_\nu A_\mu (x), \label{a8} \\
\Box A_\mu  + D^\nu D_\mu A_\nu&=&0, \qquad D_\mu A^\mu =0.
\eea{a9}
Notice that $A_\mu(x)$ does not respect the boundary condition 
$h_{\mu\nu}^{TT}(x,\pi/2)=0$, so that it cannot be decomposed as
$A_\mu=\sum_m A_\mu^m H^m$.

To bring  $h_{\mu\nu}^{TT}$ into the form given in the text, 
$h_{\mu\nu}^{TT}=\sum_m h^m_{\mu\nu}H^m$, we use a 4-d diffeomorphism:
$\epsilon_\mu=-A_\mu$. After the diffeomorphism, we are left with a tower
of massive spin-2 fields, each one carrying 5 degrees of freedom and
without any further gauge invariance. 
The role of $A_\mu$ identifies it as the St\"uckelberg field of 4-d 
diffeomorphisms.
\section*{Appendix B: a Change of Coordinates}
\renewcommand{\theequation}{B.\arabic{equation}}
\setcounter{equation}{0}
Here we exhibit an explicit change of coordinates that maps the Poincar\'e 
parametrization of $\mbox{AdS}_d$ 
into the parametrization used in the text, with slice $\mbox{AdS}_{d-1}$.

In Poincar\'e coordinates the line element of $\mbox{AdS}_d$ is
\bea
ds^2 &=& {L^2\over z^2}( dz^2 + dw^2 + d\hat{s}^2),\label{b1}
\\
d\hat{s}^2 &=& \eta_{\mu\nu} dx^\mu dx^\nu, \qquad \mu,\nu=0,..., d-3.
\eea{b2}
We are looking for a change of variables that puts the metric in the form

\beq
ds^2= \exp[2A(z)] [ dz^2 + w^{-2} (dw^2 + d\hat{s}^2)].
\eeq{b3}
We use the ansatz
\beq
z\rightarrow wz/L, \qquad w\rightarrow F(w,z).
\eeq{b4}
The metric eq.~(\ref{b1}) is transformed into 
\beq
ds^2= {L^4\over z^2 w^2} \left[\left(F_z^2 +{w^2\over L^2}\right)dz^2 +
\left(F_w^2 + {z^2\over L^2}\right)dw^2 + d\hat{s}^2\right]
\eeq{b5}
To have an $\mbox{AdS}_{d-1}$ slice we need $F_w^2 + {z^2\over L^2}=1$,
whose solution is
\beq
F(z,w) = \pm w\sqrt{1-{z^2\over L^2}} + f(z).
\eeq{b6}
The second condition we need is that the $dz^2$ term in the line element 
depends only on $z$. It gives the equation 
\beq
F_z^2 + (w/L)^2= g(z)w^2, 
\eeq{b7}
where $g(z)$ is an arbitrary function of $z$ only. 
This equation is solved by $f(z)=\,\mbox{constant}$.
At this point, the metric can be cast in the form given in eq.~(\ref{2}) with a
redefinition of $z$: $z\rightarrow h(z)$. The equation for $h$ is
\beq
h_z^2= \left(1- {h^2\over L^2}\right),
\eeq{b8}
which is solved by $h(z)=L\cos (z/L)$.

\end{document}